\documentstyle[twoside,fleqn,espcrc2,epsfig]{article}

\newcommand{\bruch}[2]{{\raisebox{0.2ex}{$#1$}/\raisebox{-0.3ex}{$#2$}}}
\newcommand{\cs}{c_{\sigma}}
\newcommand{\ct}{c_{\tau}}
\newcommand{\Ss}{S_{\sigma}}
\newcommand{\St}{S_{\tau}}
\newcommand{\Ns}{N_{\sigma}}
\newcommand{\Nt}{N_{\tau}}

\newcommand{\pa}{\partial}
\newcommand{\nn}{\nonumber}

\newcommand{\AmS}{{\protect\the\textfont2
  A\kern-.1667em\lower.5ex\hbox{M}\kern-.125emS}}

\title{Direct determination of the gauge coupling derivatives for
       the energy density in lattice QCD
        \thanks{Talk presented by T.\,Scheideler, work supported by DFG-grant
                Pe 340/3-3.}
      }
\author{F. Karsch, J. Engels and T. Scheideler
        \address{Fakult{\"a}t f{\"u}r Physik, Universit{\"a}t Bielefeld, 
                 Postfach 100131, 33501 Bielefeld, Germany}
        }
       
\begin{document}

\begin{abstract}
By matching Wilson loop ratios on anisotropic lattices we measure
the coefficients $\cs$ and $\ct$, which are required for the
calculation of the energy density. The results are compared to that of
an indirect method of determination. We find similar behaviour, the
differences are attributed to different discretization errors.
\end{abstract}

\maketitle

\section{Different ways to calculate the energy density}
From statistical mechanics we know the energy density as 
$\epsilon=\frac{1}{V} \left(\frac{\pa \ln Z}{\pa\, (1/T)}\right)_V$.
In lattice calculations the straightforward approach to calculate this
quantity is to introduce different lattice spacings $a_\sigma,~a_\tau$ to 
perform the derivatives with respect to $1/T=\Nt a_\tau$ at fixed physical
volume $(\Ns a_\sigma)^3$. Therefore one introduces the lattice anisotropy
$\xi=\bruch{a_\sigma}{a_\tau}$ and two gauge couplings in the action,
\begin{eqnarray}
  \label{action}
  S=\frac{2N}{g_\sigma^2(\xi)} \frac{1}{\xi} \sum_{\mu<\nu<4} S_{\mu,\nu}
   +\frac{2N}{g_\tau^2  (\xi)} \xi           \sum_{\mu<\nu=4} S_{\mu,4}~.
\end{eqnarray}
The energy density $\epsilon$ contains then the derivatives 
of these gauge couplings,
\begin{eqnarray}
\label{ener}
\epsilon&=&T^4 \left(\frac{\Nt}{\Ns}\right)^3\frac{2N}{g^2}
         \big( \left<\Ss-\St\right> \nn\\
        &&+  g^2\left( \cs\left< \Ss-S_0 \right>
                      +\ct\left< \St-S_0 \right> \right) \big)  ~,
\end{eqnarray}
where
{\small
\begin{tabular}{lll}
$S_0$&\multicolumn{2}{l}{\hspace*{-2ex}- action on a symmetric lattice, 
                           $T$=0 contribution,} \\
$\Ss$&\hspace*{-2ex}- spatial part of the action& \ finite temperature\\
$\St$&\hspace*{-2ex}- temporal part of the action& 
    \hspace*{-2ex}\raisebox{1.5ex}[-1.5ex]{\bigg\}}~simulation,\\
\end{tabular}
}
and
\begin{eqnarray}
  \cs\equiv \left.\frac{\pa g_\sigma^{-2}(a,\xi)}{\pa\xi} \right|_{\xi=1}
  ,~~
  \ct\equiv \left.\frac{\pa   g_\tau^{-2}(a,\xi)}{\pa\xi} \right|_{\xi=1} .
\end{eqnarray}
The coefficents $\cs$ and $\ct$ were calculated pertubatively
in \cite{karsch1}.

A different way to determine the energy density was used by the Bielefeld 
group\cite{eos}. The free energy density $f$ is obtained by integrating 
the difference of the plaquettes,
\begin{equation}
 \label{eins}
 \bruch{f}{T^4}\bigg|_{\beta_0}^\beta =
 -N_\tau^4\int_{\beta_0}^\beta d\tilde{\beta}
 \big[2S_0-(S_\sigma + S_\tau)\big]~.
\end{equation}
In large systems, where $p=-f$, $\epsilon$ can be found from the pressure,
\begin{eqnarray}
  \label{e-3p}
  \frac{\epsilon-3p}{T^4} &=& T\frac{d}{dT}\frac{p}{T^4} \\
            &=& 12N\Nt^4(\cs+\ct)\big[2S_0-(S_\sigma + S_\tau)\big] ~, \nn
\end{eqnarray}
Using the relation 
$\cs+\ct=-\frac{1}{12}\left(a\frac{\pa\beta}{\pa a}\right)$
and the $\beta$-function from $T_c/\sqrt{\sigma}$ measurements, we can
calculate $\cs$ and $\ct$ from (\ref{ener}), (\ref{eins}) and (\ref{e-3p}). 
In Figure 1 the results for $\cs$ and $\ct$ are shown
as solid lines. They deviate clearly from the pertubative results 
(broken lines). However, this is expected, as also $\cs+\ct$ deviate
from the pertubative result. We also note, that
the method works only above $\beta_{\rm critical}(N_\tau)$,
because both $p$ and $\epsilon$ become very small 
below the critical point.
\section{Direct determination of $\cs$ and $\ct$ by matching of 
         Wilson loop ratios}
An attempt to determine the lattice anisotropy was made by Burgers et 
al.\cite{burgers}. Consider the anisotropic couplings   
\begin{equation}
  g_\sigma^{-2}(\xi)=\frac{\xi}{2N}\frac{\beta}{\gamma} ~,~~  
  g_\tau^{-2}(\xi)  =\frac{1}{2N\,\xi}\,\beta\gamma~,
\end{equation}

\centerline{\epsfig{figure=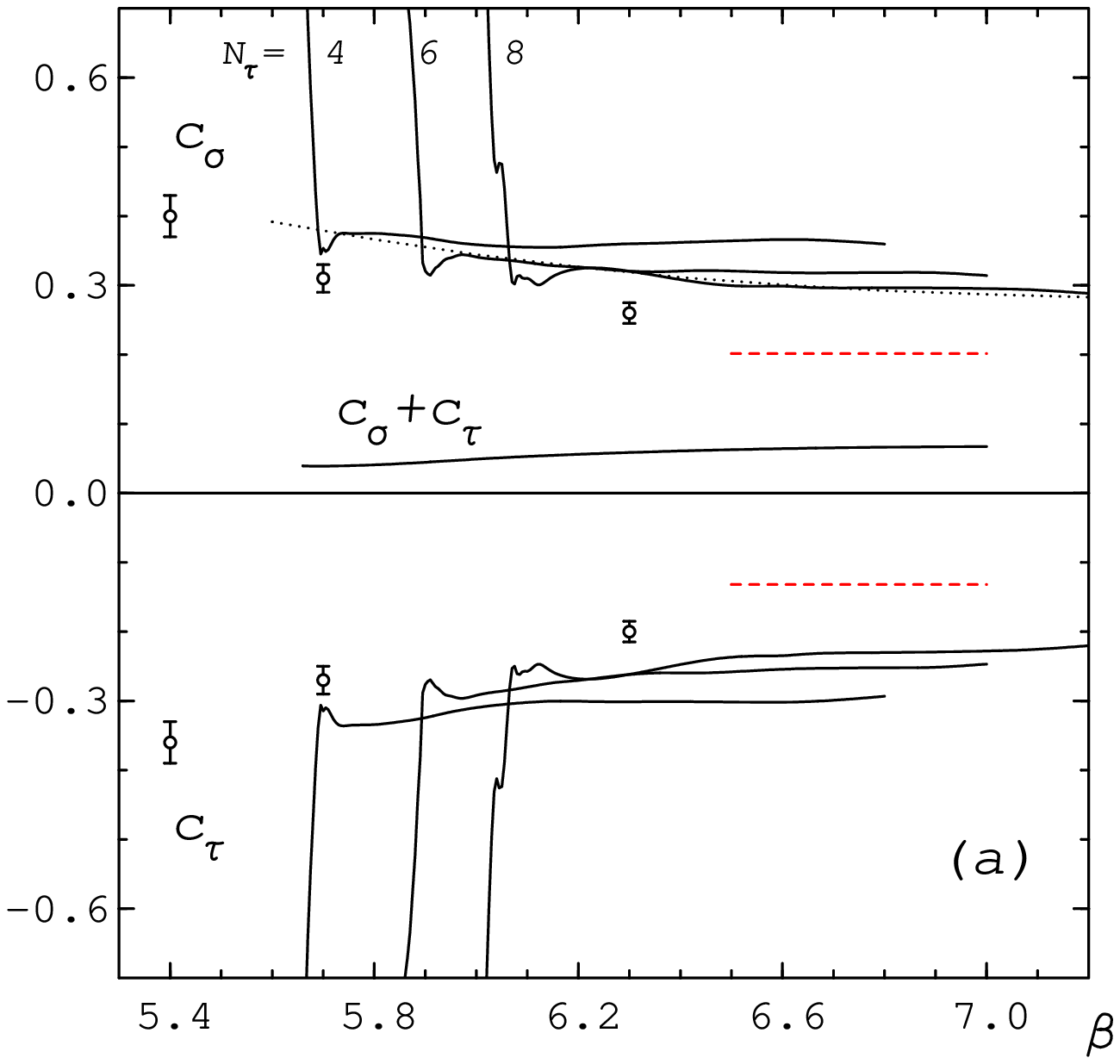,width=7.4cm}}
\vspace*{3ex}
\centerline{\hspace*{1mm}\epsfig{figure=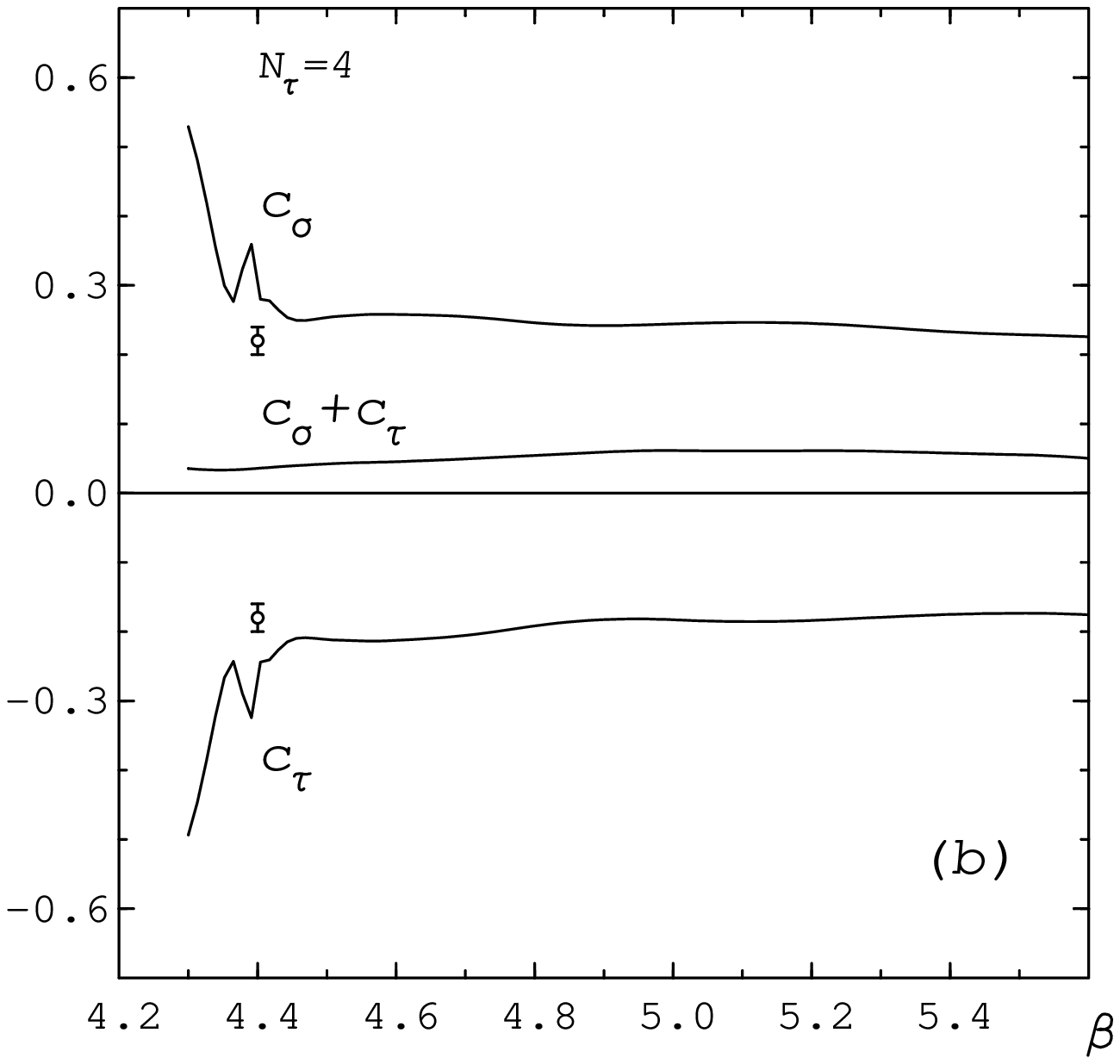,width=7.4cm}}
{\small Fig.\,1.~The derivatives $\cs$ and $\ct$ for the Wilson action
(a) ($\Nt$=4,6,8) and the 2x2-action (b) ($\Nt$=4) \cite{bibo} 
versus $\beta$. The broken lines show the pertubative values\cite{karsch1}. 
The sum $\cs+\ct$ is from 
the $\beta$-function. The dotted line shows an estimate form 
the different $\Nt$-values. The single measurements are from the matching
method.
}
\vspace*{1ex}

\noindent
of the action (\ref{action}). The derivatives with respect to the
lattice anisotropy $\xi=a_\sigma/a_\tau$ are
\begin{eqnarray}
  \label{cs}
 c_{\sigma,\tau}
  = \pm\frac{\beta}{6} \left(1-\frac{\pa\gamma}{\pa\xi}\right)-
    \frac{1}{4}\left(a\frac{\pa g^{-2}(a)}{\pa a}\right)~,
\end{eqnarray}
which requires the function $\gamma(\xi)$ or $\xi(\gamma)$.

\noindent
We vary the anisotropy of the couplings $\gamma$ in the action
$S=\bruch{\beta}{\gamma} \, S_\sigma + \beta\gamma \, S_\tau$ and 
measure the anisotropy of the lattice spacing $\xi$.
Wilson loops are suitable observables, since they depend on the
physical size of the lattice. 
Instead of matching Wilson loops, measured
in spatial and in temporal direction, directly,
\begin{equation}
  W_\sigma(x,z=\xi t) = W_\tau(x,t)~,
\end{equation}
we match ratios of Wilson loops, in order to 
cancel corner and selfmass contributions,
\begin{equation}
  R_1(x,t)=\frac{W(x,t)}{W(x+1,t)} ~~\mbox{or}
\end{equation}
\begin{equation}
  R_2(x,t)=\frac{W(x+1,t)W(x-1,t)}{W(x,t)^2}~.
\end{equation}
Since these ratios contain only Wilson loops with the same extension in $t$,
the following matching condition holds,
\begin{equation}
  R_\sigma(x,z=\xi t) = R_\tau(x,t)~.
\end{equation}

\begin{center}
\epsfig{figure=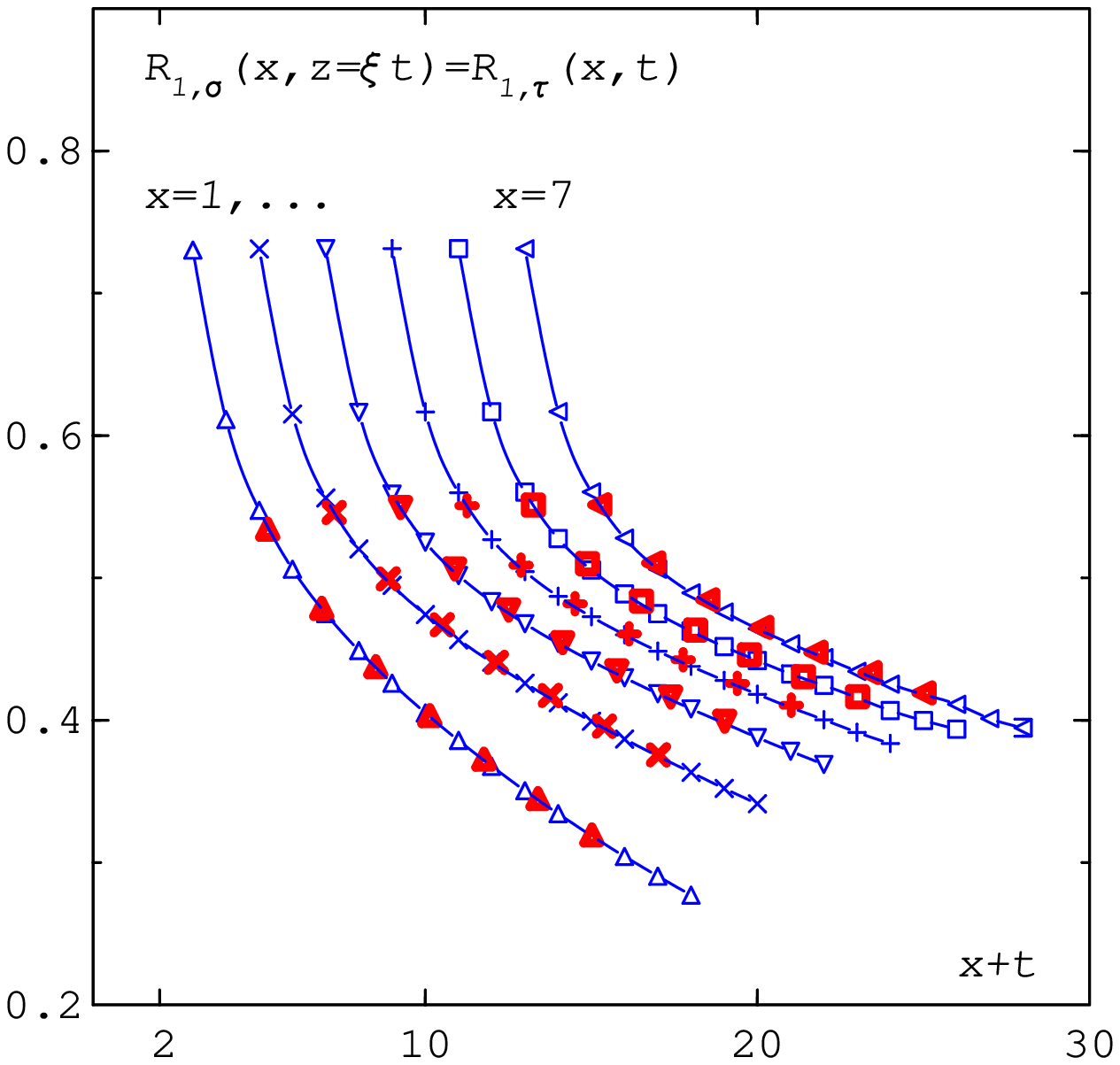,width=7.4cm}
\end{center}
{\small  Fig.\,2.~The ratios $R_{1,\tau}(x,t)$ for fixed $x=1,\dots,7$ vs.
$t+x$, the open symbols are connected by solid lines. The ratios 
$R_{1,\sigma}(x,z)$ correspond to the bold face symbols and are shifted
in $t$ by $\xi=1.63(4)$.
}
\vspace*{1ex}

We are using a $16^4$ lattice for $\gamma\in[0.92,1.08]$,
a $16^3\times32$ lattice for $\gamma\in[1.1,2.0]$ and 
a $16^3\times48$ lattice for $\gamma=3.0$. The link integration
technique of ref.\,\cite{pdf} is applied to obtain accurate expectation values 
of large Wilson loops with a statistics of 2000 measurements.

The value of $\xi$ was chosen in such a 
way, that the square deviation of $R_{1,\sigma}(x,z=\xi t)$ from the
lines connecting the $R_{1,\tau}(x,t)$ measurements
was minimal. For each $\gamma$ we find in this way a value
of $\xi$. Figure 3 shows an example for $\beta=6.3$. In all cases 
we found a linear behaviour of $\xi(\gamma)$.

\begin{center}
\epsfig{figure=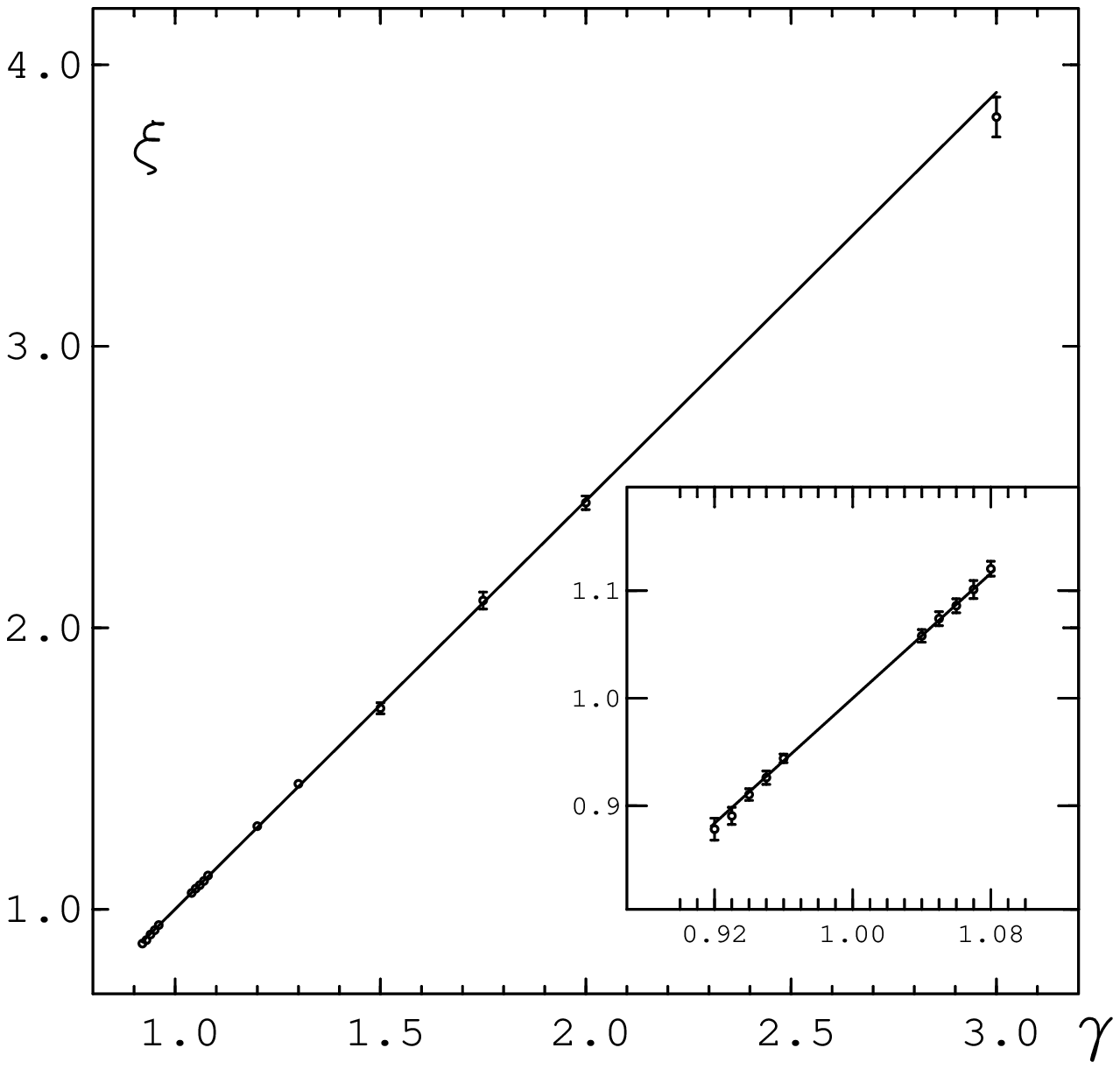,width=7.4cm}
\end{center}
{\small Fig.\,3.~The result for $\xi$ as a function of $\gamma$ 
for $\beta=6.3$. The solid line is a linear fit to $\xi(\gamma)$.
}
\vspace*{1ex}

To obtain the derivative $\left.\pa\xi/\pa\gamma\right|_{\xi=1}$ we have
performed linear fits. The resulting $\cs$ are shown in Figure 4.

Obviously the matching procedure requires at least loops of an area eight.
For larger Wilson loops the results are consistent.

\section{Summary and conclusion}
Employing a method of matching ratios of Wilson loops on anisotropic lattices
we have directly determined the derivatives $\cs$ and $\ct$ of the
gauge couplings non-pertubatively. This has been done for both, the standard 
Wilson action and the 2$\times$2 improved action\cite{bibo}. 
We found significant deviations from the pertubative result. 
Qualitatively our results are in agreement with the indirect method used
in\cite{eos}, though they are a little lower. We assume that the difference
is due to ${\cal O}(a^n)$ corrections, since they enter the two methods
differently.

The use of ratios of Wilson loops was crucial in order to eliminate
unphysical self energy contributions. The numerical
matching technique for these ratios enabled us to obtain the 
required accuracy for $\xi$. 

\begin{center}
\epsfig{figure=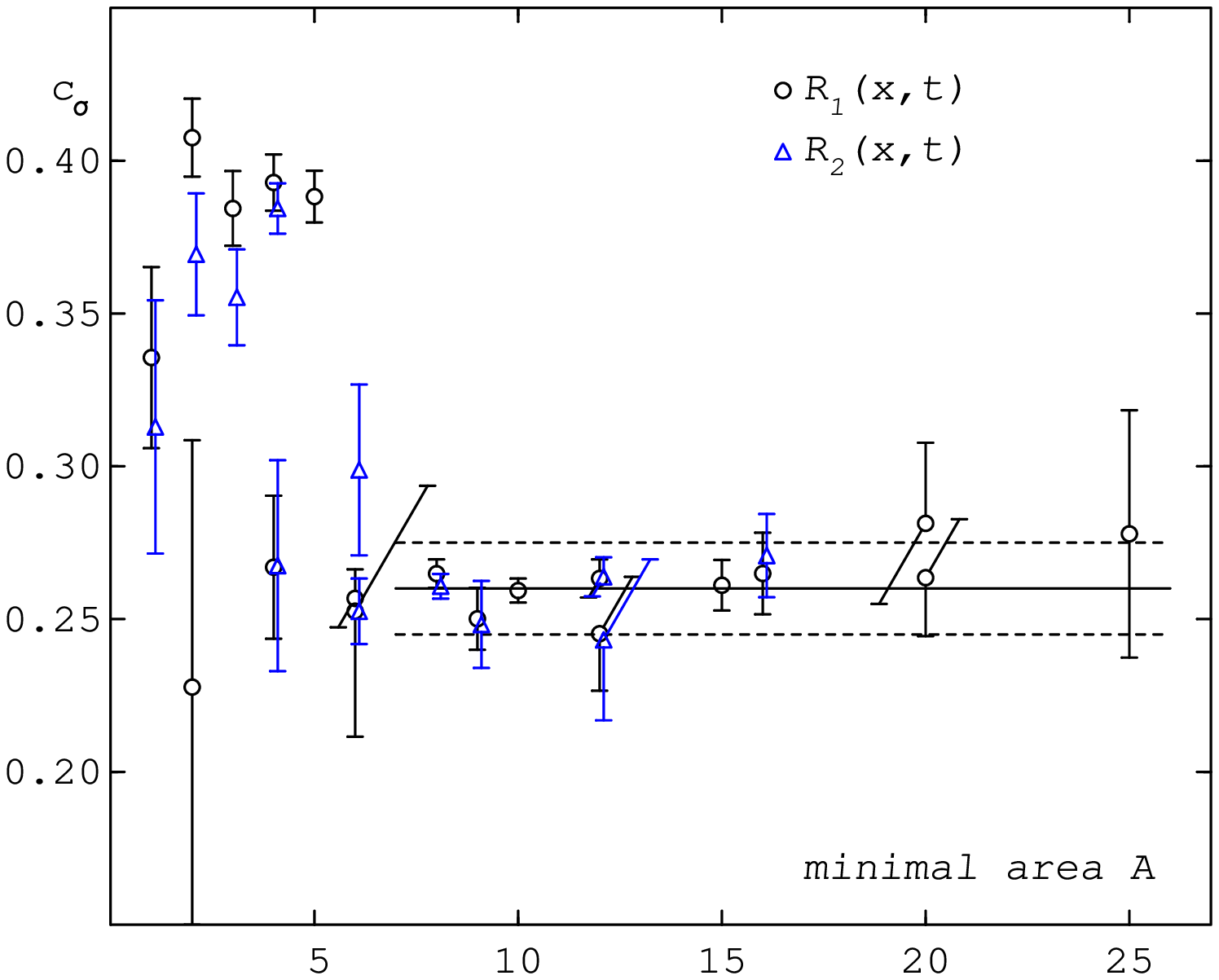,width=7.4cm}
\end{center}
{\small Fig.\,4.~The derivative $\cs$ for $\beta=6.3$ from Eq.(\ref{cs}) as
a function of the minimal area $A$ of the smallest Wilson loops included
in the ratio matching. The solid line shows the average value
for $A\ge8$, the broken lines the error band.
}

\end{document}